\documentclass[10pt,conference]{IEEEtran}
\usepackage{amssymb}
\usepackage{latexsym}
\usepackage{graphicx}

\begin{document}

\newcommand{\calc}{{\cal C}}
\newcommand{\F}{{\mathbb F}}
\renewcommand{\vec}[1]{{\mathbf #1}}
\newcommand{\ip}[1]{{\langle #1 \rangle}}
\newcommand{\tup}[1]{{\left( #1 \right)}}
\newcommand{\FOR}{{\bf for}}
\newcommand{\mytab}{\hspace*{1.5em}}
\newcommand{\sem}[1]{{\sf{\em #1}}}
\newcommand{\Z}{{\mathbb Z}}
\newcommand{\R}{{\mathbb R}}
\newcommand{\kwd}[1]{\ {\bf \small #1}\ }
\newtheorem{definition}{Definition}[section]

\title{ Approximate MAP Decoding on Tail-Biting Trellises}
\author{\authorblockN{Madhu A.S.}
\authorblockA{Department of Computer Science \& Automation.\\
Indian Institute of Science\\
Bangalore, India. \\
Email: madhu@csa.iisc.ernet.in}
\and
\authorblockN{Priti Shankar}
\authorblockA{Department of Computer Science \& Automation.\\
Indian Institute of Science\\
Bangalore, India. \\
Email: priti@csa.iisc.ernet.in}
}
\maketitle
\begin{abstract}
We propose two approximate algorithms for MAP decoding on tail-biting trellises. The algorithms work 
on a subset of nodes of the tail-biting trellis, judiciously selected. 
We report the results of simulations on an AWGN channel using the approximate algorithms on tail-biting trellises for the $(24,12)$ Extended Golay Code
 and a rate $1/2$ convolutional code with memory $6$.  
\end{abstract}
\section{Introduction}
In this paper we propose a new approximate MAP decoding technique on tail-biting trellises
 that exploits the subtrellis 
structure of the tail-biting trellis to compute approximate aposteriori probabilities (APPs) of codeword symbols.
Our algorithm is best described as a best-search
 algorithm, meaning thereby, that the algorithm tries to track those subtrellises 
which are most likely to contain the transmitted codeword and computes marginals 
over these only. Our approximate decoder
works rather efficiently at higher values of signal to 
noise ratios. We compare our results with those obtained by the {\em wrap}
version of Anderson-Hladik MAP decoder~\cite{anderson}, which we refer 
to as the AH-decoder.
The rest of the paper is organized as follows.
Section~\ref{sec:background} gives some background.
 Section~\ref{sec:decoding} describes the decoding algorithm. 
Section~\ref{sec:modification} describes a simple modification to the scheme.
Section~\ref{sec:simulations} presents the results of simulations on an AWGN
 channel on a $16$ state tail-biting trellis for the $(24,12)$ Extended Golay 
code and a $64$ state tail-biting trellis for a rate $1/2$, memory $6$,  
convolutional code. 
Section~\ref{sec:conclusion}
 concludes the paper.
\section{Background}
\label{sec:background}
Tail-biting trellises for convolutional codes were introduced in~\cite{solomon} and those for block codes in~\cite{cald,kv1}.
For ease of notation we view the tail-biting trellis $T=(V,E,{\F}_q)$, of depth
$n$, to be defined on a
sequential time-axis with $V_0=V_n$ and all valid paths restricted to
those that begin and end at the same state. We call $V_0$ and $V_n$
respectively the start and final states of $T$.
In~\cite{sh1,sb,shao,solomon} it was shown that a tail-biting trellis could be viewed as the
superposition of subtrellises obtained from a coset decomposition of the code represented by it with respect to a subgroup.
Corresponding to each start state of the tail-biting trellis we define a subcode consisting of
all codewords that begin and end at the same state. These subcodes all have
identically structured subtrellises (though with different labels), and share 
states at various time indices. This subtrellis structure is exploited to give an approximate MAP decoding algorithm.
\section{Approximate Decoding on Tail-biting Trellises}
\label{sec:decoding}
In order to explain the approximate MAP algorithm 
we define the weight of a subtrellis $T_j$ conditioned on the received vector $\vec{r}$ as 
\[
weight(T_j)=\sum_{\vec{c} \in \calc_j} Pr(\vec{c}|\vec{r})
\]
where $C_j$ is the subcode represented by $T_j$.
The weight of a subtrellis is the aposteriori
probability that the transmitted codeword lies in the 
particular subtrellis. 
The weight of a subtrellis is also the maximum contribution a subtrellis
can make to the decision sums for computing the APP of any digit of the 
codeword. 
The basic idea in MAP Approximate Algorithm (MAA) is to 
start off by obtaining initial estimates to the weights of the subtrellises. 
The initial estimates are all overestimates. We begin with the best subtrellis with 
respect to the current estimates and begin a forward pass on that subtrellis. 
We define our updating function for estimates as we progress along sections of the current trellis 
such that the overestimates become more and more accurate as we go along, always converging to the 
exact value at the end. Each time we move from one section to the next, we check the updated 
estimate against those of other subtrellises and switch to another subtrellis if it appears more 
likely to be the correct one. Thus the algorithm may switch from one subtrellis to another during the
course of execution. However, it is always guaranteed to finish on the most likely subtrellis. 
The same is done for the backward pass.
Now after the two passes are over, there will be subtrellises for which certain sections have been opened 
only in the forward or backward pass but not both. In the final marginalization phase we ignore these sections of
subtrellises which have not been opened by both the passes.
\subsection{Node Objective Functions}
\label{subsec:objective}
Given a one-to-one tail-biting trellis $T = (V, E, \F_q)$ of depth $n$, we use the following notation.
We denote the label of an edge $e \in T$ by
$l(e)$. For an edge $e=(u,v) \in T$  we define predecessor and successor operators as
$Init(e)=u$ and $Term(e)=v$. By nodes of $T$ we refer to the vertices of $T$.
The start node of a subtrellis $T_j$ is denoted by $s_j$ and the final
node by $f_j$.
The set of all paths in the tail-biting trellis from node $u$ to  node $v$ is denoted by $P(u,v)$. 
Further let $P(U,V)=\cup_{u\in U,v\in V}P(u,v)$ denote the set of paths from nodes in 
$U$ to nodes in $V$.

Given a received vector $\vec{r}$ we annotate the edges of the tail-biting
trellis suitably using the channel information so that the 
codeword APP gets decomposed along the edges of the codeword path. 
The weight of an edge is denoted by $w(e)$. 
 We then define the weight of a path as the product of the weights of the edges constituting the path.
The weight of a subtrellis $T_j$, as defined previously, is then the
sum of the weights of all paths in $T_j$. 

The approximate algorithm computes a set of
node-objective functions on the nodes of the tail-biting trellis. These
function definitions are identical to the forward-backward passes defined
in~\cite{bcjr}.

$\alpha_{T}$ and $\beta_{T}$ are functions whose domain is the nodes of the
tail-biting trellis $T$.
They are defined recursively as
\begin{eqnarray} \label{eqn:1}
\alpha_{T}(v)&:=&\sum_{e:\mathit{Term}(e)=v} \alpha_{T}(\mathit{Init}(e)) w(e) \nonumber \\
\beta_{T}(v)&:=&\sum_{e:\mathit{Init}(e)=v} \beta_{T}(\mathit{Term}(e)) w(e)\end{eqnarray}
with $\alpha_{T}(V_0):=\beta_{T}(V_n):=1$.\\
This is just a forward-backward pass on $T$, initialized according to the boundary conditions.

Also corresponding to each subtrellis $T_j$ we define two functions 
$\alpha_{T_j}$ and $\beta_{T_j}$ whose domain is the set of nodes 
belonging to $T_j$. $\alpha_{T_j}$  and $\beta _{T_j}$ at a node $u$ capture the computational
effects of exclusive forward and backward passes respectively on $T_j$ at $u$. 

$\alpha_{T_j}$ at a node $u \in T_j$ is defined inductively as
\begin{eqnarray} \label {eqn:2}
\alpha_{T_j}(u)&:=&\sum_{e \in T_j:\mathit{Term}(e)=u} \alpha_{T_j}(\mathit{Init}(e)) w(e)
\end{eqnarray}
with $\alpha_{T_j}(s_j):=1$. \\
Similarly $\beta_{T_j}$ at a node $u \in T_j$ is defined inductively as
\begin{eqnarray}\label{eqn:3}
\beta_{T_j}(u)&:=&\sum_{e \in T_j:\mathit{Init}(e)=u} \beta_{T_j}(\mathit{Term}(e))w(e)
\end{eqnarray}
with $\beta_{T_j}(f_j):=1$.\\
It can be seen that $\alpha_{T_j}$ at a node $u \in T_j$ 
is the sum of the weights of all paths in $P(s_j,u)$, while 
$\beta_{T_j}$ at $u$ is the sum of the weights of all paths in $P(u,f_j)$.
Similarly $\alpha_T$ at a node $u \in T$ gives the sum of the weights of all paths in $P(V_0,u)$ whereas
$\beta_{T}$ at $u$ gives the sum of the weights of all paths in $P(u,V_n)$.

\subsection{The Approximate MAP Algorithm}
\label{subsec:approx}
We now give an informal description of the MAP Approximate Algorithm (MAA).\\

\noindent
{\sf MAP-AA} \\
{\tt Input:} A one-to-one tail-biting trellis  $T = (V, E, \F_q)$ of depth $n$, with edge
weights suitably defined using the received vector $\vec{r}$ and channel information \\
{\tt Output:} A vector of approximations to $Pr(c_i=\sigma|\vec{r})$ for $\sigma \in \F_q$
and $i=1,2,\ldots,n$ 
\subsubsection{Phase 1}
This phase computes the node-objective functions $\alpha_T$ and $\beta_T$ 
with respect to the tail-biting trellis $T$ by executing a forward-backward pass on $T$ with boundary conditions as suggested by Recursion~\ref{eqn:1}.
\subsubsection {Phase 2}
This phase computes a set of node-objective functions $\alpha_{T_j}$
and $\beta_{T_j}$
with respect to the subtrellises that share the node. It consists
of a forward and a backward pass.

A node can be shared among many subtrellises
and will belong to the domain of the functions defined with respect to these
subtrellises. 
MAA computes a subset of these functions either partially or completely.
\paragraph{Forward Pass}
At each step in the forward pass the approximate algorithm first chooses a 
winning subtrellis $T_w$. It then computes $\alpha_{T_w}$ for the next section
of $T_w$ using Recursion~\ref{eqn:2}. 
The winning subtrellis is one at which a suitably defined heuristic
function, $h_f$, is maximized. 
We associate a working index with each subtrellis. The working index
of a subtrellis $T_j$ gives the last section of the tail-biting trellis 
at which the node-objective function $\alpha_{T_j}$ has been computed.

The forward pass works along the following lines.
The working indices are initialized to the start section and the boundary
conditions of Recursion~\ref{eqn:2} are enforced. 
We start by choosing a winning subtrellis $T_w$ from the set of
$|V_0|$ subtrellises defined on $T$ such that $h_f$ is maximized at $T_w$. 
The heuristic function
at a subtrellis $T_j$ with a working index $k$ is a function of $\alpha_{T_j}$ and $\beta_{T}$ 
at section $k$ in $T$.
We describe the heuristic function in detail later. If the working index of $T_w$ is the final section, we successfully exit 
from the forward pass. Otherwise we increment the working index to the next section. 
Using the values of $\alpha_{T_w}$ at the nodes of previous working index of $T_w$,
we then compute $\alpha_{T_w}$ at the nodes of the current working index according to
Recursion~\ref{eqn:2}.
After computing $\alpha_{T_w}$ for the
current working index,we re-evaluate the heuristic function at $T_w$.
We then go back to the process of choosing the winning subtrellis and 
computing the corresponding  node-objective function at the next working index of the winning subtrellis. 

The heuristic function at $T_j$ with working index $k$ is defined
as
\[
h_f(T_j,k):=\sum_{u \in V_k\cap T_j}\alpha_{T_j}(u)\beta_{T}(u).\]
We now motivate this definition of the heuristic function.\\
Let $\Pi_{j}^k$ be the set of paths in $P(V_0,V_n)$ whose first $k$ edges lie in  $T_j$.
It can be seen that $h_f(T_j,k)$ is the sum of the weights of  paths in
$\Pi_{j}^k$. Note that $h_f(T_j,0)=\beta_T(s_j)$ is the sum of weights of
paths in $P(s_j,V_n)$.\\
Now observe that $\Pi_{j}^{k+1}\subseteq \Pi_{j}^k$.
Since the edge-weights are probabilities and therefore non-negative, this
implies that \[h_f(T_j,k+1)\leq h_f(T_j,k).\]
Also by definition $\Pi_{j}^n=T_j$ and as a consequence
$h_f(T_j,n)=weight(T_j)$.\\
It follows that the heuristic for $T_j$ keeps falling after each revision and finally converges to the weight
of $T_j$. Thus at any instant the heuristic for $T_j$ is an over-estimate to the weight of $T_j$.
If a forward pass has been completed on a subtrellis $T_j$, 
the approximate algorithm guarantees that a forward pass will be completed on 
all subtrellises $T_k$ with $weight(T_k) > weight(T_j)$.

\paragraph{Backward Pass}
In the backward pass we compute a set of node-objective functions
$\beta_{T_j}$ with respect to subtrellises $T_j$ as dictated by Recursion~\ref{eqn:3}. 
The backward pass is similar
in spirit to the forward pass except for the definition of the heuristic
function and the backward direction of computational flow on the trellis.
The computation starts by initializing the working indices of subtrellises to
the final section and ends when the working index of the winning subtrellis
is the start section.

The heuristic function at $T_j$ with working index $k$ for the backward pass
is defined as
\[
h_b(T_j,k):=\sum_{u \in V_k \cap T_j}\beta_{T_j}(u)\alpha_{T}(u).
\]
It can be seen that $h_b(T_j,k)$ is the sum of weights of all paths in $P(V_0,V_n)$ 
whose last $n-k$ edges lie completely in $T_j$.
It follows that all the properties noted for the heuristic function along the
forward pass carry over to the heuristic function for the backward pass.

\subsubsection{Phase 3}
This phase computes the approximate marginals corresponding to each symbol 
$\sigma$ and each position $i$.
We compute the approximate aposteriori probabilities 
$Pr(c_i=\sigma|\vec{r})$ as
\[\sum_{e \in E_i,l(e)=\sigma}w(e)
\sum_j\alpha_{T_j}(Init(e))\beta_{T_j}(Term(e)).
\]
The product in the inner-sum is taken over only those node-objective
functions which have been computed.

\section{A Simple Modification}
\label{sec:modification}
By  restricting the Phase 2 of MAA to 
work with a fixed number of subtrellises say $\mu$, we can reduce the storage requirements of the
algorithm at the cost of incurring a further penalty in the accuracy of the APPs computed. 
In order to decide the $\mu$ subtrellises to work with, we evaluate
$\mbox{min}(h_f(T_j,0),h_b(T_j,n))$ at each subtrellis $T_j$ and choose the
first $\mu$ subtrellises at which this quantity is the largest.\\
The rationale behind this choice is that both $h_f(T_j,0)$ and $h_b(T_j,n)$ are
overestimates to the $weight(T_j)$ and the minimum of the two is nearer to the
true weight of $T_j$.
We call this modified scheme as the $\mu\mbox{-MAP Approximate Algorithm}$ ($\mu\mbox{-MAA}$). 
Surprisingly this scheme gives pretty good results for the codes on which we have run experiments. 
\begin{figure}
	\centering
	\includegraphics[width=2.5in]{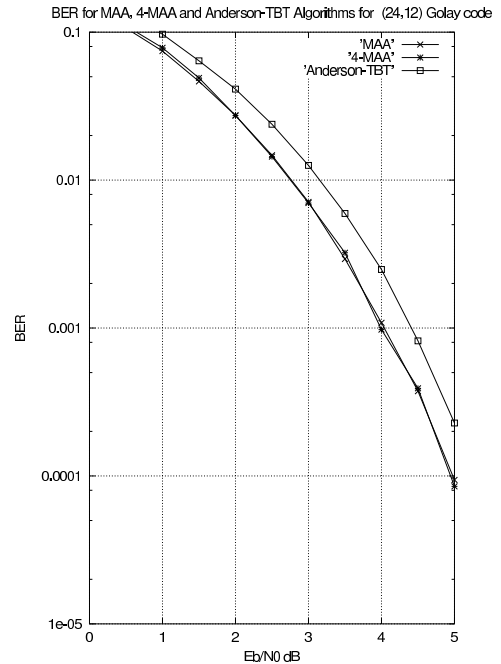}
	\caption{BER for the MAA, 4-MAA and AH-Decoder (Wrap=10) algorithm for the (24,12) Extended Binary Golay Code}
	\label{fig:golay-ber}
\end{figure}

\begin{figure}
	\centering
	\includegraphics[width=2.5in]{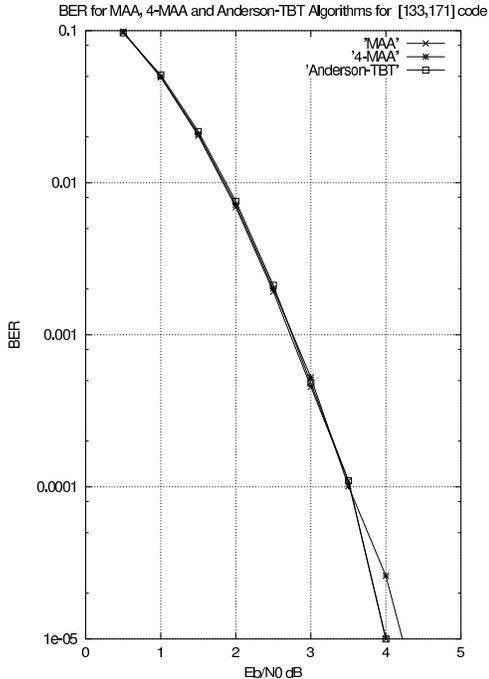}
	\caption{BER for the MAA, 4-MAA and AH-Decoder (Wrap=40) algorithm for the rate 1/2 (133,171) Convolutional Code with circle length 48}
	\label{fig:conv1-ber}
\end{figure}
\section {Simulation results}
\label{sec:simulations}
We have coded the MAA, 4-MAA and AH-Decoder and shown the results of simulations on the minimal 
16 state tail-biting trellis~\cite{cald} for the extended (24,12,8) Golay code 
and a 64 state tail-biting trellis for a rate 1/2 convolutional codes with memory 6 (equivalent to the (554,744) convolutional code of~\cite{anderson}), with circle size 48. This is the same code experimented on in~\cite{anderson} .
The channel model used is an AWGN channel with antipodal signaling. The source bits were assumed to
be equally-likely.

For the convolutional code we show the variation of the average number of forward-backward updates
with the signal to noise ratio for the MAA and compare it with the number of forward-backward updates
required by the AH-Decoder.\\ 
We also show the variation of average number of subtrellises explored by
 the Phase 2 of MAA with signal to noise ratio. The tail-biting trellis representing the
 code has $3072$ states and $64$ subtrellises. Each subtrellis has $2493$ states.
 The result is displayed in Table~\ref{tab:expansions2}. 
It can be seen that at moderate to high SNR, the Phase 2 of MAA seems to work on only a single
subtrellis.

We also display the performance of the MAA, 4-MAA and AH-Decoder in Figures~\ref{fig:golay-ber} and~\ref{fig:conv1-ber} and  find that there is virtually no difference in the bit error rates for the three algorithms
for the convolutional code. For the Golay code the MAA and the 4-MAA seem to do 
slightly better than the AH-Decoder.  
\begin{table}[!ht]
\begin{center}
\begin {tabular} {|c|c|c|c|}
\hline
{\small \bf SNR}& {\tiny \bf avg updates }& {\tiny \bf avg updates by }& {\tiny \bf avg no: of subtrellises}\\
&{\tiny \bf by MAA} & {\tiny \bf AH-Decoder} & {\tiny \bf examined by MAA}\\
\hline
0.0&    91867&  22528  &7.60 \\
\hline
0.5&    53737 &22528   &4.04 \\
\hline
1.0&    34113 &22528   &2.16 \\
\hline
1.5&    25984	&22528   &1.38 \\
\hline
2.0&    23087  &22528   &1.11 \\
\hline
2.5&    22230  &22528   &1.02 \\
\hline
3.0&    22049  &22528   &1.00 \\
\hline
3.5&    22014  &22528   &1.00 \\
\hline
4.0&    22008  &22528   &1.00 \\
\hline
4.5&    22008  &22528   &1.00 \\
\hline
5.0&    22008  &22528   &1.00 \\
\hline
\end{tabular}
\caption{\small Runtime statistics for the MAA and AH-Decoder (Wrap=40) for the rate $1/2$, memory $6$, $[133,171]$ convolutional code with circle 
length $48$. 
\label{tab:expansions2}}
\end{center}
\end{table}
\section{Conclusion}
\label{sec:conclusion}
We have shown that at the expense of some extra space we can obtain approximate algorithms with good performance for MAP decoding on tail-biting trellises. Simulations on tail-biting trellises for the (24,12) Extended Golay code and a 
rate $1/2$, memory $6$ convolutional code used in~\cite{anderson} have been carried out and the results on an AWGN channel are reported. 
\\
\\
\\

\end{document}